\documentclass[conference]{IEEEtran}
\IEEEoverridecommandlockouts
\usepackage{cite}
\usepackage{amsmath,amssymb,amsfonts}
\usepackage{algorithmic}
\usepackage{graphicx}
\usepackage{textcomp}
\usepackage{xcolor}
\usepackage{comment}

\def\BibTeX{{\rm B\kern-.05em{\sc i\kern-.025em b}\kern-.08em
    T\kern-.1667em\lower.7ex\hbox{E}\kern-.125emX}}

\usepackage[font=footnotesize]{subcaption}
\usepackage[font=footnotesize]{caption}

\newif\ifaccepted
\acceptedtrue 

\begin{document}
\newcommand\redbf[1]{\textcolor{red}{\textbf{#1}}}

\title{Walsh-domain Neural Network for Power Amplifier Behavioral Modelling and Digital Predistortion}

\ifaccepted
\author{\IEEEauthorblockN{Cel Thys\IEEEauthorrefmark{1}\IEEEauthorrefmark{3}, Rodney Martinez Alonso\IEEEauthorrefmark{1}, Antoine Lhomel\IEEEauthorrefmark{2}, Maxandre Fellmann\IEEEauthorrefmark{2},\\ Nathalie Deltimple\IEEEauthorrefmark{2}, Francois Rivet\IEEEauthorrefmark{2}, Sofie Pollin\IEEEauthorrefmark{1}}
\IEEEauthorblockA{\IEEEauthorrefmark{1}WaveCoRE, Department of Electrical Engineering (ESAT), Katholieke Universiteit Leuven, Belgium}
\IEEEauthorblockA{\IEEEauthorrefmark{1}Univ. Bordeaux, CNRS, Bordeaux INP, IMS, UMR 5218, F-33400 Talence, France}
\IEEEauthorblockA{\IEEEauthorrefmark{3}email: cel.thys@kuleuven.be}
}
\fi
\maketitle

\begin{abstract}
This paper investigates the use of Neural Network (NN) nonlinear modelling for Power Amplifier (PA) linearization in the Walsh-Hadamard transceiver architecture. This novel architecture has recently been proposed for ultra-high bandwidth systems to reduce the transceiver power consumption by extensive parallelization of the digital baseband hardware. The parallelization is achieved by replacing two-dimensional quadrature modulation with multi-dimensional Walsh-Hadamard modulation. The open research question for this architecture is whether conventional baseband signal processing algorithms can be similarly parallelized while retaining their performance. A key baseband algorithm, digital predistortion using NN models for PA linearization, will be adapted to the parallel Walsh architecture. A straighforward parallelization of the state-of-the-art NN architecture is extended with a cross-domain Knowledge Distillation pre-training method to achieve linearization performance on par with the quadrature implementation. This result paves the way for the entire baseband processing chain to be adapted into ultra-high bandwidth, low-power Walsh transceivers.
\end{abstract}


\section{Introduction}
Future wireless networks should evolve beyond human-centric wideband demand, to serve billions of devices simultaneously, while concurrently supporting high-throughput machine-to-machine communication applications like chip-to-chip ultra-high-speed communications~\cite{rappaport_subTHz_2019}. However, spectrum availability and management have become the main bottleneck for the future development of wireless networks. These factors have pushed the development of one of the key enabling technologies for 6G and beyond, ultra-wideband sub-terahertz (sub-THz) communications. 

One of the main challenges to enable THz communications is that the transceivers are more affected by non-linearities across the ultra-wideband frequency range~\cite{TeraCom1}. For instance, for an ultra-wideband system with 10~GHz of bandwidth, high-speed data converters with a Nyquist sampling rate higher than 20~Giga-samples-per-second (GSPS) are required. However, state-of-the-art converters for ultrawideband applications underperform in terms of their Walden and Schreier figures of merit ~\cite{murmann_adcsurvey,BookMurmann2020}. These figures of merit respectively consider the trade-off between sampling rate, power consumption, and effective number of bits, and between sampling rate, power consumption and the signal-to-noise and distortion ratio. Therefore, the converter´s linearity and energy efficiency for ultrawideband wireless communications is quite limited~\cite{manganaro2022ADCintro}.

To overcome the hardware limitations imposed by the large sampling rate requirements for ultrawideband communications in near-THz spectrum, the Walsh-Hadamard transceiver architecture has been recently proposed~\cite{bouassida_walshTX_2016, ferrer2023walshAWG}. This transceiver architecture splits the ultrawideband communication system into multiple parallel undersampling digital chains. Each chain modulates a distinct Walsh-Hadamard basis function with a certain \textit{sequency}, similar to OFDM subcarriers with their distinct frequencies. Since each chain has a low digital sampling rate, the overall system power consumption is much lower than an equal bandwidth quadrature system \cite{ferrer2023walshAWG}. The architecture can be used with any arbitrary waveform, using Walsh-Hadamard series approximation~\cite{bouassida_walshTX_2016}. The Walsh-Hadamard transformation allows mapping time-domain symbols into a multidimensional orthogonal constellation space. Therefore, an \textit{N}-order Walsh transformation will enable an \textit{N}-dimensional constellation mapping and the same number of parallel processing chains in the receiver, reducing each ADC's sampling rate by a factor \textit{N}~\cite{WalshRx}.

Although the Walsh transformation allows relaxing the requirements in terms of energy, linearity and performance of the analog-to-digital converters, it requires entirely new implementations of baseband algorithms, such as digital predistortion (DPD) to compensate power amplifier non-linearities. Indeed, high bandwidth systems require the use of wideband power amplifiers (PA), which have an inherent tradeoff between power efficiency and linearity. The operation region with higher input power will be more power efficient but also exhibit more nonlinear behavior. To improve power efficiency without sacrificing linearity, many have proposed digital predistortion linearization techniques \cite{eun_indirect_1997, morgan_generalized_2006, fellmann_walshlms, brihuega_frequency-domain_2021, liu_dynamic_2004,wang_augmented_2019, wu_residual_2020, DNNvsVolterra, hongyo_2019_DNNDPD}. In DPD, the communication signal is sent through an inverse behavioral model of the PA (the predistorter) before going to the PA itself, such that the predistorter cancels out the PA's nonlinear distortion.

One class of DPD algorithms uses the linear-in-parameters Volterra series \cite{eun_indirect_1997, morgan_generalized_2006, fellmann_walshlms, brihuega_frequency-domain_2021} to implement the predistorter model. The parameters of these models can be identified using a single least-squares fit or iteratively using algorithms such as recursive least squares or least mean squares \cite{eun_indirect_1997, fellmann_walshlms}. Parallelizing Volterra-based DPD has also been investigated, for instance in~\cite{brihuega_frequency-domain_2021} authors propose a separate DPD model for each subcarrier in an OFDM system.  A Volterra predistorter model in \textit{N}-dimensional parallel Walsh domain has been presented for the first time by authors in~\cite{fellmann_walshlms}. 

One of the challenges faced by traditional Volterra predistorters is that some predistorted waveforms can have a larger PAPR compared to the signal itself, particularly for OFDM systems~\cite{DNNvsVolterra}. Another limitation of Volterra models is the high correlation between polynomial bases of higher orders, thus limiting the scaling of Volterra models to high model complexities \cite{hongyo_2019_DNNDPD}.
To overcome these challenges, deep neural network models have been widely investigated for DPD \cite{liu_dynamic_2004, wang_augmented_2019, wu_residual_2020, DNNvsVolterra, hongyo_2019_DNNDPD}. 

The basic neural network (NN) model used for PA behavioral modelling and distortion is the Real-Valued Time-Delay Neural Network (RVTDNN) \cite{liu_dynamic_2004}. In this model, the complex baseband signal is split into real and imaginary components which are fed to a feedforward NN, together with a vector of previous samples. The outputs of the NN are the real and imaginary components of the predistorted waveform, to be sent to the RF frontend. The NN model parameters are identified iteratively using methods based on stochastic gradient descent.
The performance of the basic RVTDNN was improved in \cite{wang_augmented_2019}, augmenting it with additional inputs containing envelope terms (e.g. amplitude or amplitude squared).
Authors in~\cite{DNNvsVolterra} have investigated a type of TDNN with delayed and advanced input samples, and found that it reduces the PAPR of the predistorted waveform compared to Volterra predistorters.
Wu et al. \cite{wu_residual_2020} incorporated a skip connection between inputs and outputs of the predistorter neural network, this forces the neural network layers to model additive nonlinearity. They named the resulting model the Residual Real-valued TDNN (R2TDNN). 

To the best of our knowledge, the adaptation of R2TDNN models to N-dimensional parallel Walsh domain and the tradeoff between the performance indicators of the DPD and the neural network complexity has not been investigated before. Indeed, our preliminary research showed that a DPD based on R2TDNN extended to N-dimensional Walsh domain is not capable of fully linearizing the response of ultrawideband power amplifiers. Although this approach improves the EVM, the spectral regrowth of the PA is not significantly reduced. We hypothesize that during the training for $N$ dimensions, the neural network optimizer loses track of essential features of the time-frequency response of the amplifier as the amplitude and phase components are transformed into a higher dimensional space (i.e., Walsh domain). We propose pre-training of the Walsh-domain neural network predistorter, using the predistortion waveform generated by a large IQ-domain model. This process is known as Knowledge Distillation \cite{hinton2015distilling}, wherein the IQ-DPD model is the teacher and the Walsh-DPD model the student that learns to mimic the teacher's output. 
Multiple authors~\cite{Distillation, fischer_phasenorm} have similarly used offline pre-training with an ideal predistortion waveform acquired using computationally expensive iterative learning control \cite{chani_cahuana_2016_ILC}.

The main novelty of this paper is an adaptation of R2TDNN DPD models for an N-dimensional parallel Walsh-Hadamard communication system, without sacrificing the linearization performance. More specifically, the R2TDNN model will be adapted to the Walsh domain by replacing the IQ inputs and outputs with N-dimensional vectors, similar to the process followed in \cite{fellmann_walshlms} for Volterra models. To precondition the Walsh-domain predistorter towards specific features relevant to time domain DPD, we propose Knowledge Distillation offline pre-training with an IQ-domain DPD acting as teacher model. 

The outline of this paper continues as follows. In Section~\ref{sec:methods} we detail the proposed neural network for PA behavioral modeling and DPD in the Walsh sequency domain. In Section~\ref{sec:results} we provide our research findings and provide a numeric analysis demonstrating that the forward modelling and linearization performance of the proposed approach is on par with quadrature-based R2TDNN. Finally, our research findings and conclusions are summarized in section \ref{sec:conclusion}.

\section{Proposed Walsh-domain Neural Network} \label{sec:methods}

\begin{figure}[h]
\centerline{\includegraphics[width=0.9\linewidth]{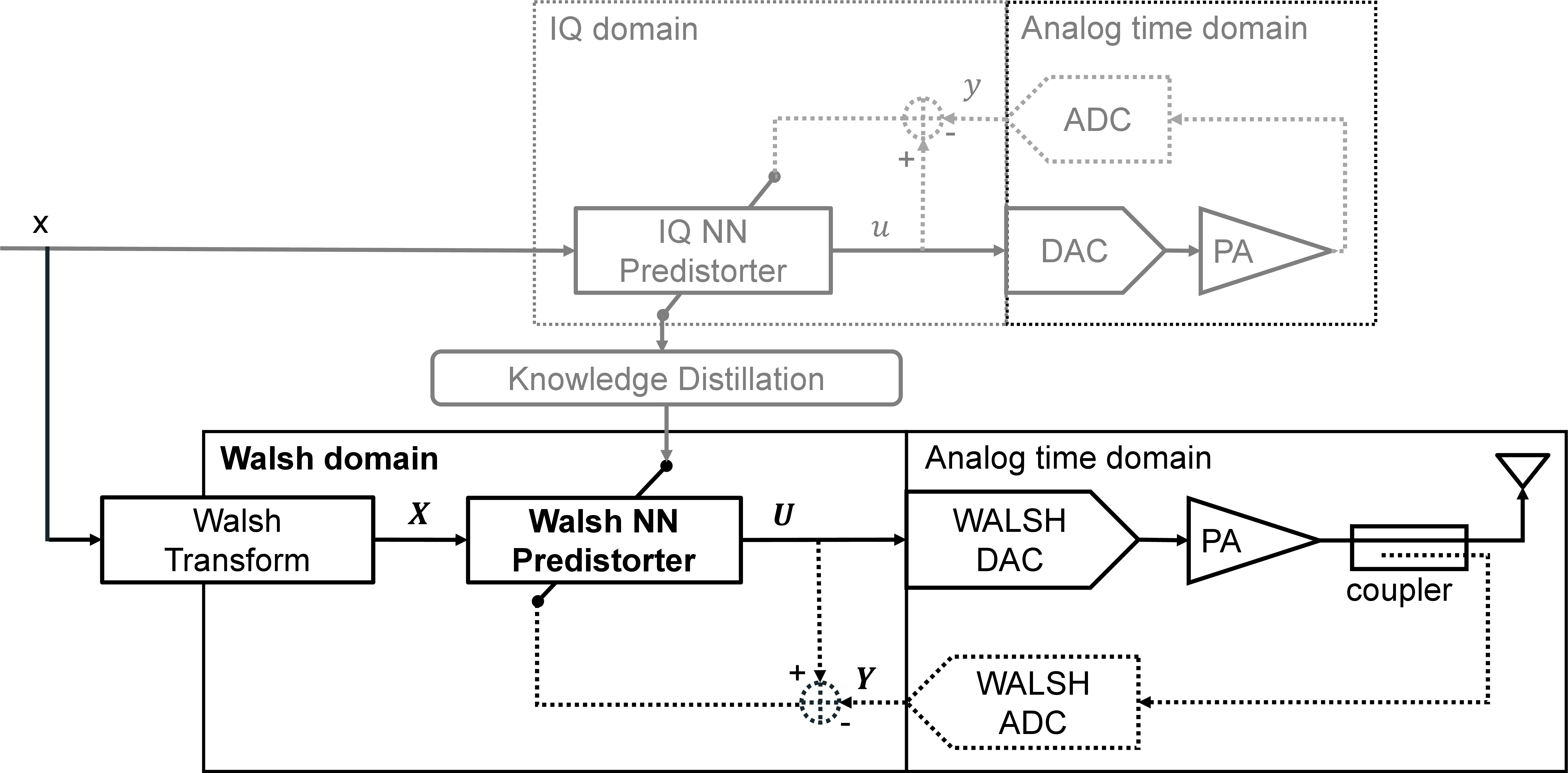}}
\caption{Rationale of proposed Walsh-based Neural Network system enhanced by off-line knowledge distillation pre-training}
\label{fig:walshdpd}
\end{figure}

The proposed approach to DPD linearization of PAs in Walsh domain is shown in Fig.~\ref{fig:walshdpd}.

In this diagram, complex IQ signals are shown in lower case $x$, while Walsh-domain vectors are shown in bold upper case $\textbf{X}$. The transmit IQ waveform $x(n)$ is transformed to Walsh-domain, predistorted using a Neural Network model, and then sent towards the PA. The IQ-domain predistorter, a R2TDNN, is drawn in grey since it is only used for offline pre-training of the Walsh-domain predistorter. This process called cross-domain Knowledge Distillation (KD) uses an ideal predistortion waveform generated by the R2TDNN as pre-training data for the Walsh-domain Neural Network (WDNN). This allows the WDNN predistorter to model accurately the ideal predistortion behavior, without specific adaptations in the model architecture. 

Whereas conventional DPD operates on IQ signals, the proposed DPD operates on Walsh-domain vectors $\textbf{X}$, which can be obtained from the respective IQ signal $x$ by means of the Walsh transform
\begin{equation}
\textbf{X}(i) = \sum_{n=1}^{N} {x(n) W^N_i(n)} \text{ for  $i=1...N$}\label{eq:walshdecomp}
\end{equation}
where $W^N_i(n)$ represents the Walsh-Hadamard basis function with Walsh order $N$ and sequency $i$ \cite{bouassida_walshTX_2016, fellmann_walshlms}. These basis functions are binary, orthogonal waveforms that decompose the signal $x(n)$ into it's sequency spectrum $\textbf{X}(i)$. By letting the predistorter operate in the sequency domain, we parallelize the DPD operation by a factor equal to the Walsh order $N$. The algorithm proposed in \cite{fellmann_walshlms} exploits this architecture to adapt a block-based Volterra model as predistorter, while our work will investigate the use of a WDNN predistorter. 

\begin{figure}[htbp]
\centerline{\includegraphics[width=0.9\linewidth]{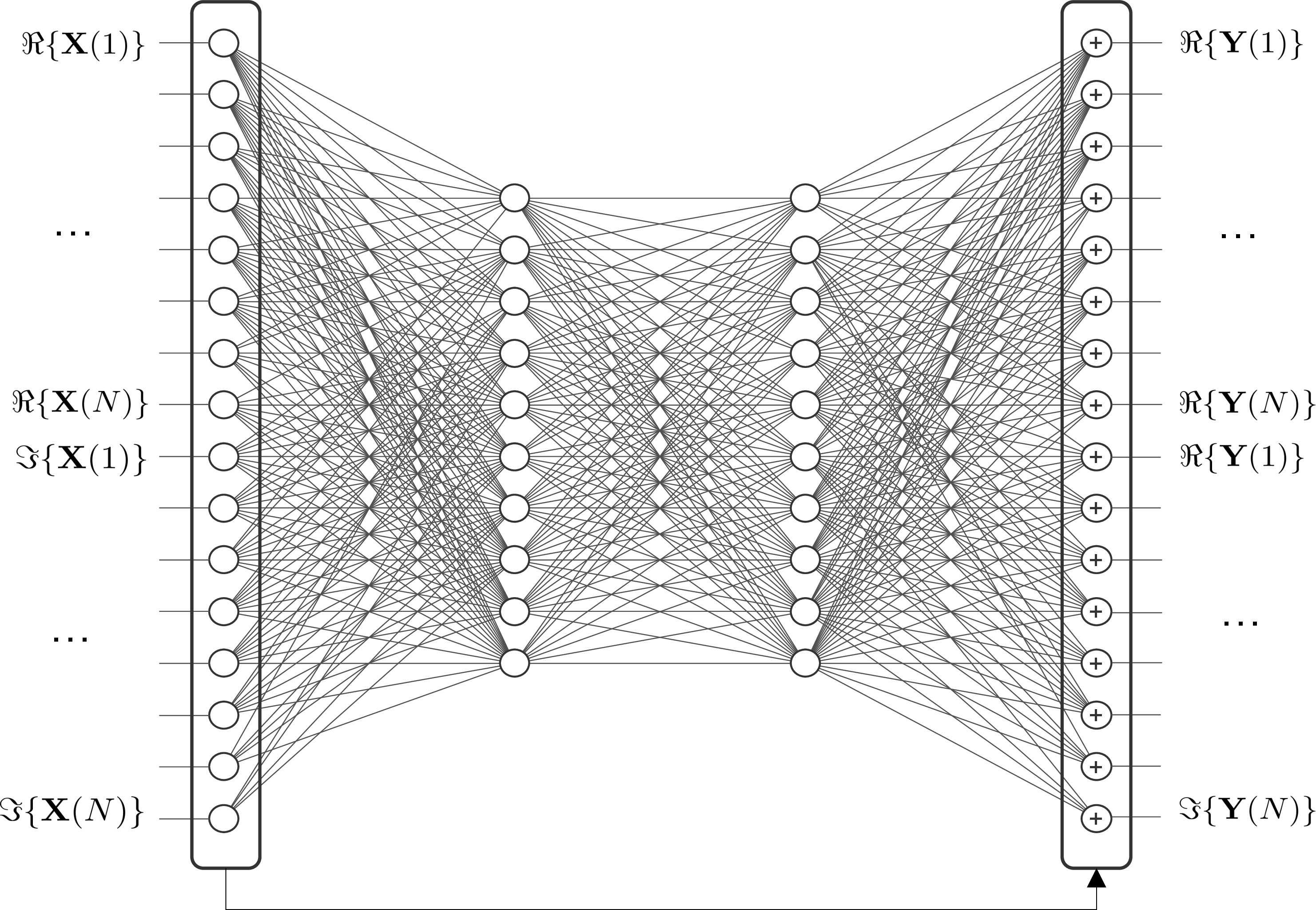}}
\caption{Proposed Walsh-domain Neural Network}
\label{fig:walshnn}
\end{figure}

In conventional neural network predistorters, the model architecture is specifically engineered for PA behavorial modelling. This includes a delay line at the NN input containing memory terms \cite{liu_dynamic_2004}, further input features representing envelope terms $|x(n)|^k$ \cite{wang_augmented_2019} and possibly a skip connection between input and output \cite{wu_residual_2020}. In our proposed WDDN architecture, the input delay line is replaced with a vector of Walsh coefficients representing the Walsh transform of the complex input. The conventional I and Q outputs are replaced by a vector output of size $2*N$, corresponding with the sequency domain version of I and Q outputs. The complete model is shown in Fig.~\ref{fig:walshnn}, including a residual connection between input and output. Although no PA-specific envelope or memory term features are input to the model, it can still function as an accurate PA behavioral model due to our Knowledge Distillation pre-training method.

The neural network behavioral model is trained using the Adam optimizer\cite{kingma2017adam}, normalized mean square error (NMSE) loss \eqref{eq:nmse} and maximum 10000 epochs of training. Training data is grouped into training-validation-test sets according to a 60-25-15\% split. Hyperparameters of the neural network model are selected using a grid search, where training is stopped early if after 50 epochs there is no improvement in validation loss. We investigated neural networks with hyperbolic tangent and ReLU activations, and similar to \cite{hongyo_2019_DNNDPD} we consistently found the tanh outperformed the ReLU activation. 

By setting a constraint on the overall number of network weights, we can search for optimal models within a certain model complexity range. We use as model complexity metric the number of Floating-Point-Operations-per-Second (FLOPS) where $k$, $n$, $f_{symb}$, $I$, $O$ represent the number of hidden layers, amount of neurons per layers, symbol rate and Input / Output feature size respectively:
\begin{equation}
    FLOPS = \left[2*n*(I + k*n + O) + O \right] * f_{symb} . \label{eq:flops}
\end{equation}

In this equation $f_{symb}$ is 20 Giga-symbols-per-second for IQ-based R2TDNN (1GHz signal bandwidth, 10 times oversampled). Walsh-based WDNN will have an $N$ times larger $I$ and $O$, but also $N$ times smaller $f_{symb}$, so overall similar FLOP numbers as IQ-based neural networks.

The NMSE is used as loss function for NN training as well as the metric to evaluate forward PA modelling accuracy on the test dataset. In \eqref{eq:nmse}, $y(n)$ represents the true PA output and $\hat{y}(n)$ represents the NN model estimated output:

\begin{equation}
   NMSE_{dB} = 10 \cdot log_{10} \left\{ \frac{\frac{1}{N}\sum\limits_{n=1}^{N}|\hat{y}(n)-y(n)|^2}{\frac{1}{N}\sum\limits_{n=1}^{N}|y(n)|^2} \right\} . \label{eq:nmse}
\end{equation}

To compare NN based predistorters, we use the error vector magnitude (EVM) and adjacent channel leakage ratio (ACLR) as performance metrics. Definitions are given below in \eqref{eq:evm} and \eqref{eq:aclr}, where $x(n)$ is the original transmit signal, $y(n)$ is the output of the combined DPD-PA system and $P_{adj}$, $P_{main}$ represent adjacent channel and main channel radiated power respectively:
\begin{equation}
EVM_\% = 100*\sqrt{\frac{\frac{1}{N}\sum\limits_{n=1}^{N}|y(n)-x(n)|^2}{\frac{1}{N}\sum\limits_{n=1}^{N}|x(n)|^2}} , \label{eq:evm}
\end{equation}

\begin{equation}
ACLR = 10*log10(\frac{P_{adj}}{P_{main}}) . \label{eq:aclr}
\end{equation}

For the PA and DPD modelling we considered a waveform with 1~GHz bandwidth and 11.49dB PAPR as input stimulus to a custom-designed CMOS PA operating in D-band. The resulting output waveform is obtained using Cadence post-layout simulation, including the PA memory effects. The custom PA in question has 11~dBm saturation power and 15~dB nominal gain.

\section{Results} \label{sec:results}
In this section we compare the Walsh-based WDNN and IQ-based R2TDNN architectures, with the Walsh-domain results always reported using Walsh order $N=64$. First, we evaluate forward PA modelling accuracy in terms of NMSE.
\begin{figure}[htbp]
\centerline{\includegraphics[width=0.9\linewidth]{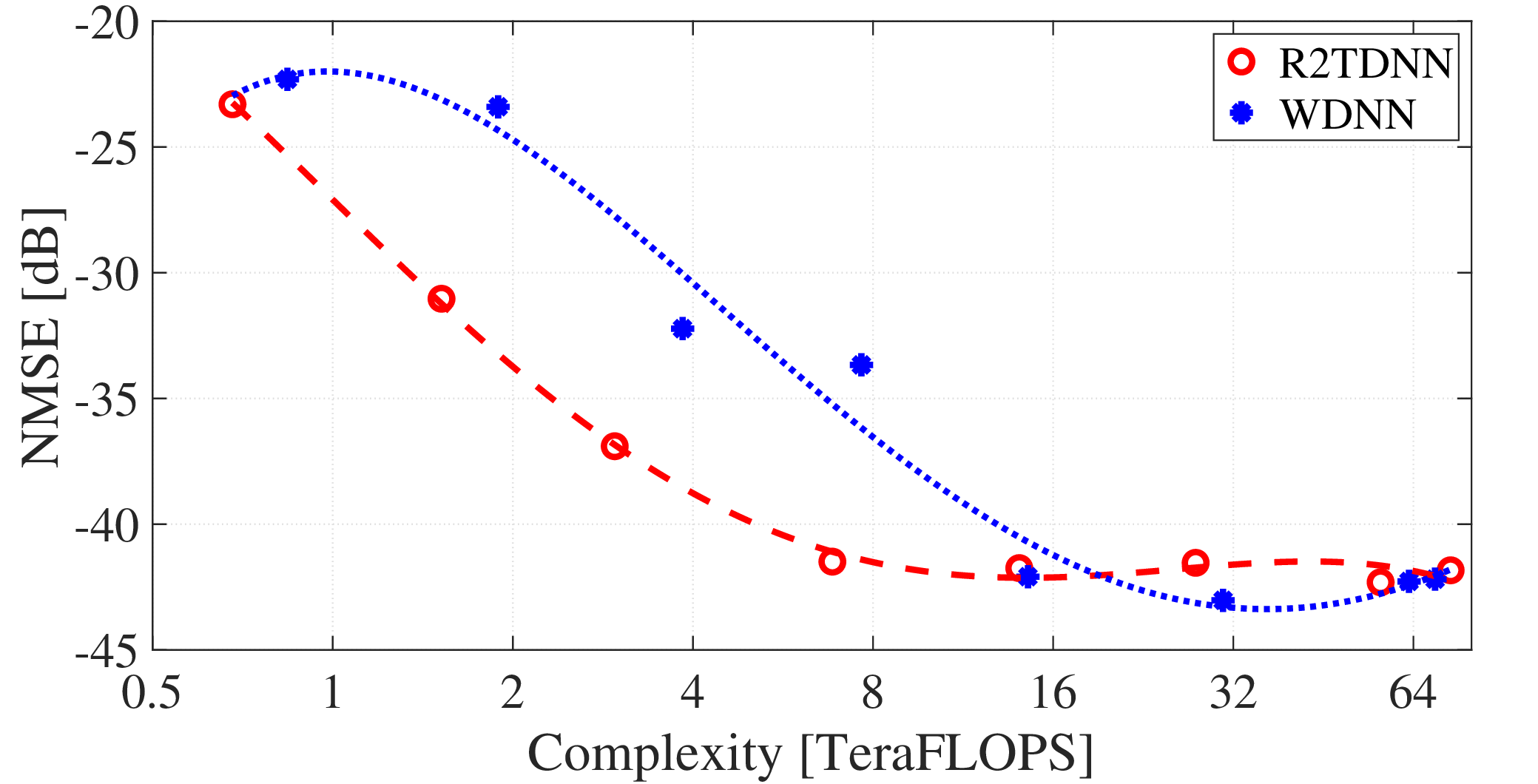}}
\caption{PA modelling accuracy (NMSE) dependence on model complexity.}
\label{fig:pamodelling}
\vspace{-4mm} 
\end{figure}
The accuracy/complexity tradeoff in Fig.~\ref{fig:pamodelling}, shows the scaling trend for both models is very similar. Indeed, the difference in accuracy between Walsh and IQ models from 16~TFLOPS onwards is negligible. For lower TFLOP numbers there is a performance gap due to the reduced capacity of the WDNN to model essential time-domain features. 

\begin{figure}[htbp]
\centerline{\includegraphics[width=0.95\linewidth]{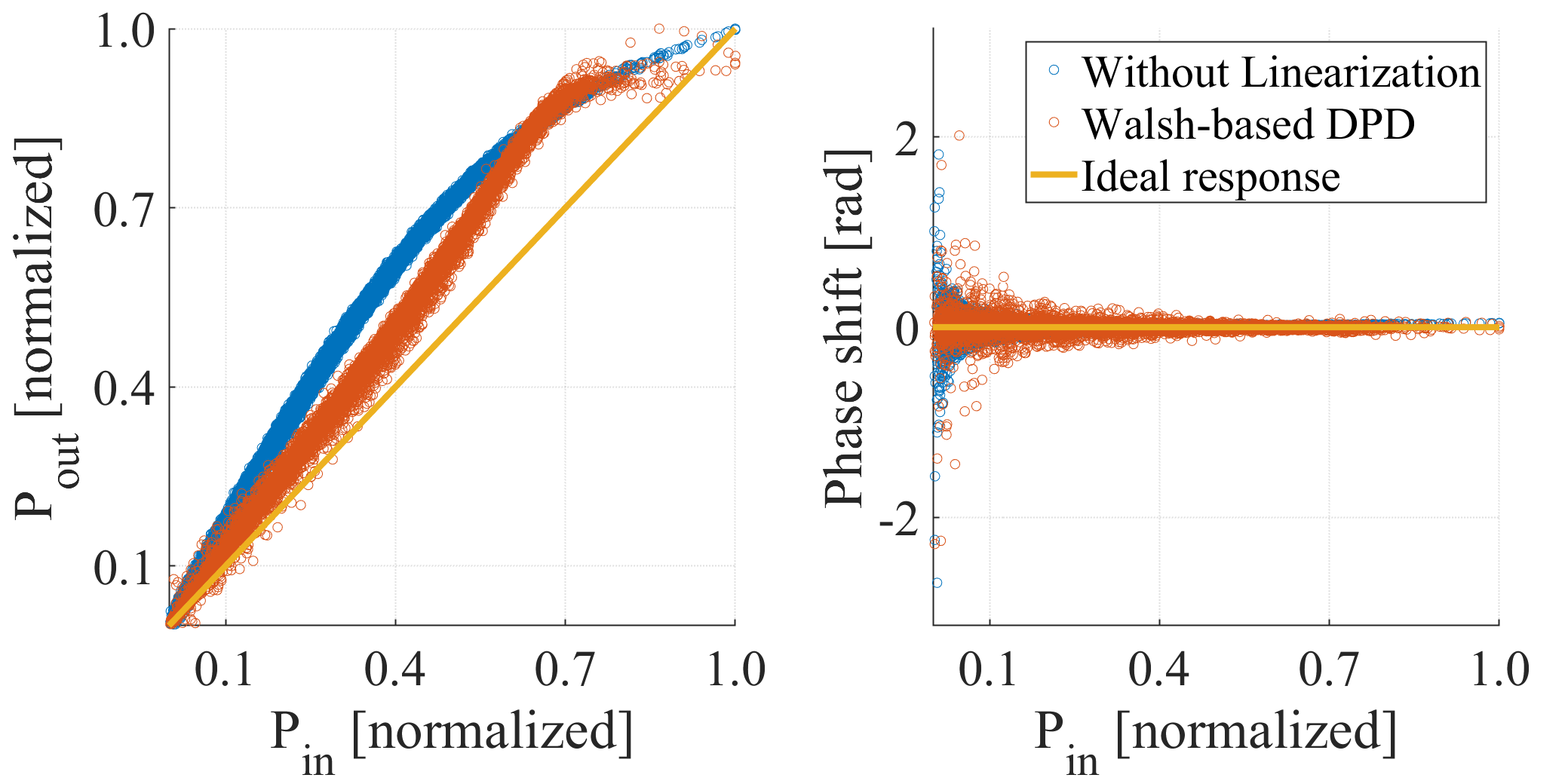}}
\caption{AM-AM and AM-PM response after applying proposed Walsh-based DPD algorithm}
\label{fig:amamdpd}
\vspace{-4mm} 
\end{figure}

\begin{figure}[htbp]
\centerline{\includegraphics[width=0.95\linewidth]{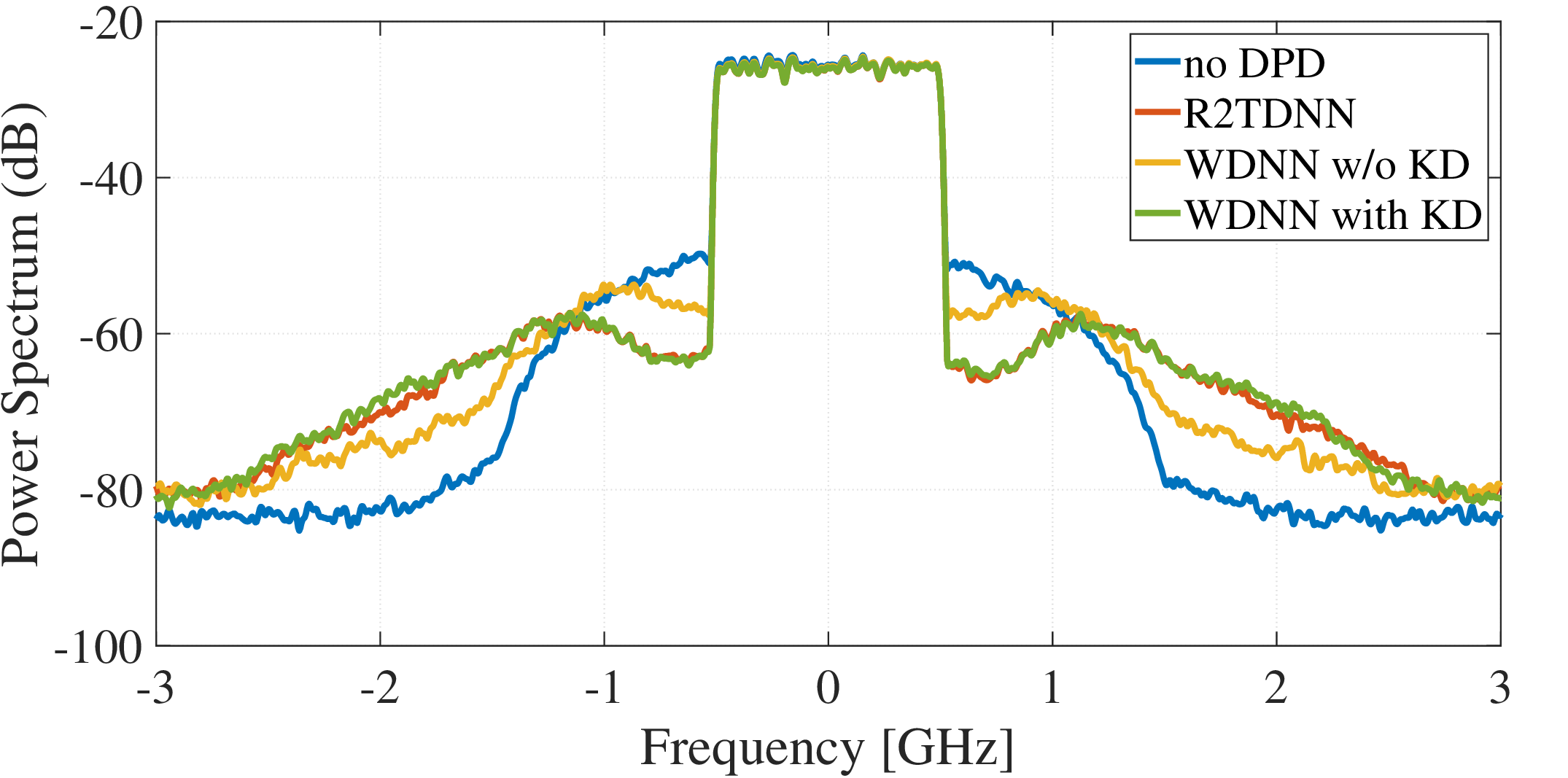}}
\caption{Spectrum at output of PA after applying considered DPD approaches}
\label{fig:spectrumdpd}
\vspace{-2mm} 
\end{figure}

For WDNN-based DPD this issue will be solved by using cross-domain Knowledge Distillation (KD) pre-training, which we validate for the simulated 1~GHz bandwidth waveform. Fig.~\ref{fig:amamdpd} shows the simulated amplitude and phase response of the linearized PA, while Fig.~\ref{fig:spectrumdpd} shows the spectrum before and after DPD.

\begin{figure}[htbp]
\centerline{\includegraphics[width=0.95\linewidth]{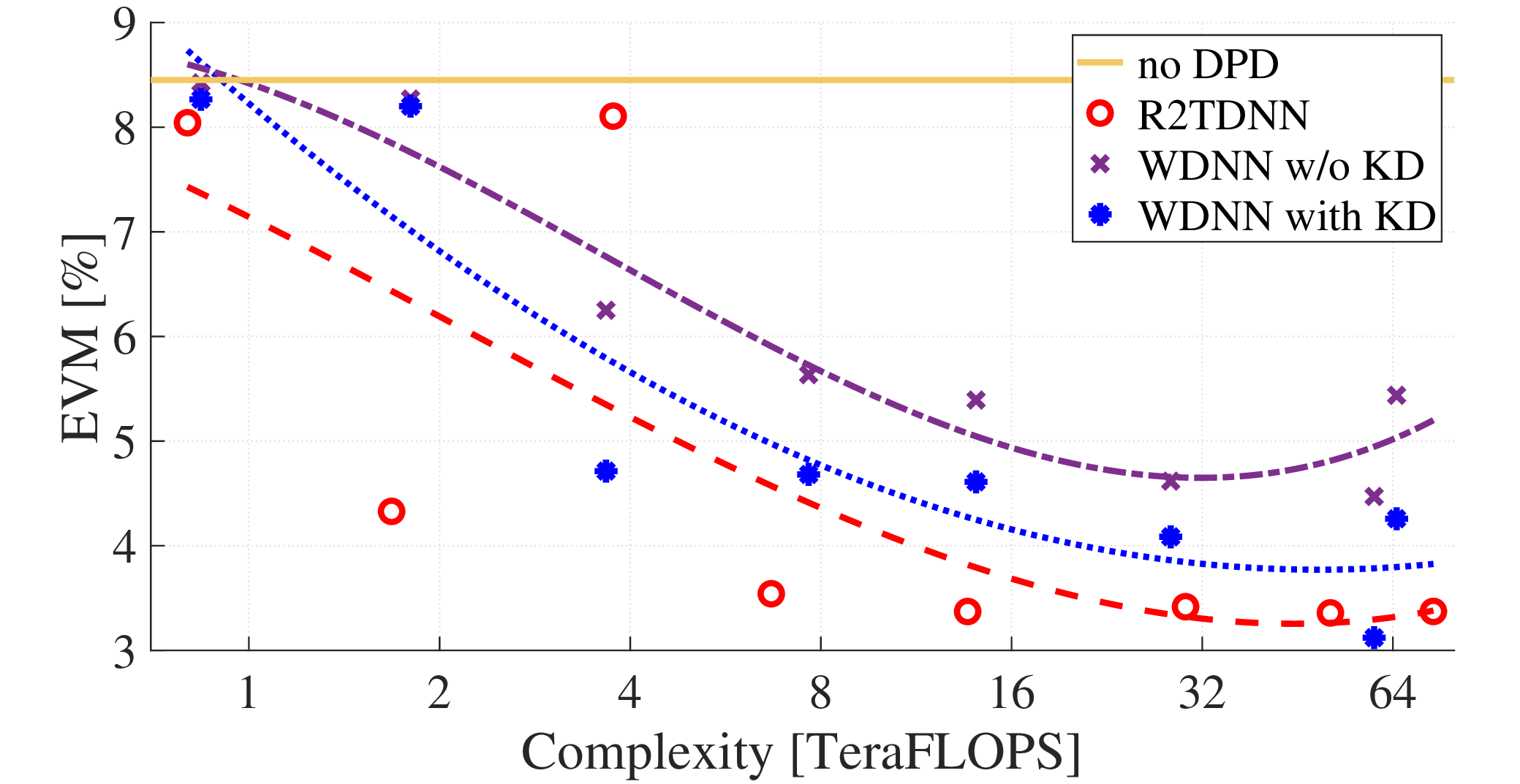}}
\caption{EVM - complexity tradeoff for investigated DPD methods}
\label{fig:evmtradeoff}
\end{figure}

\begin{figure}[htbp]
\centerline{\includegraphics[width=0.95\linewidth]{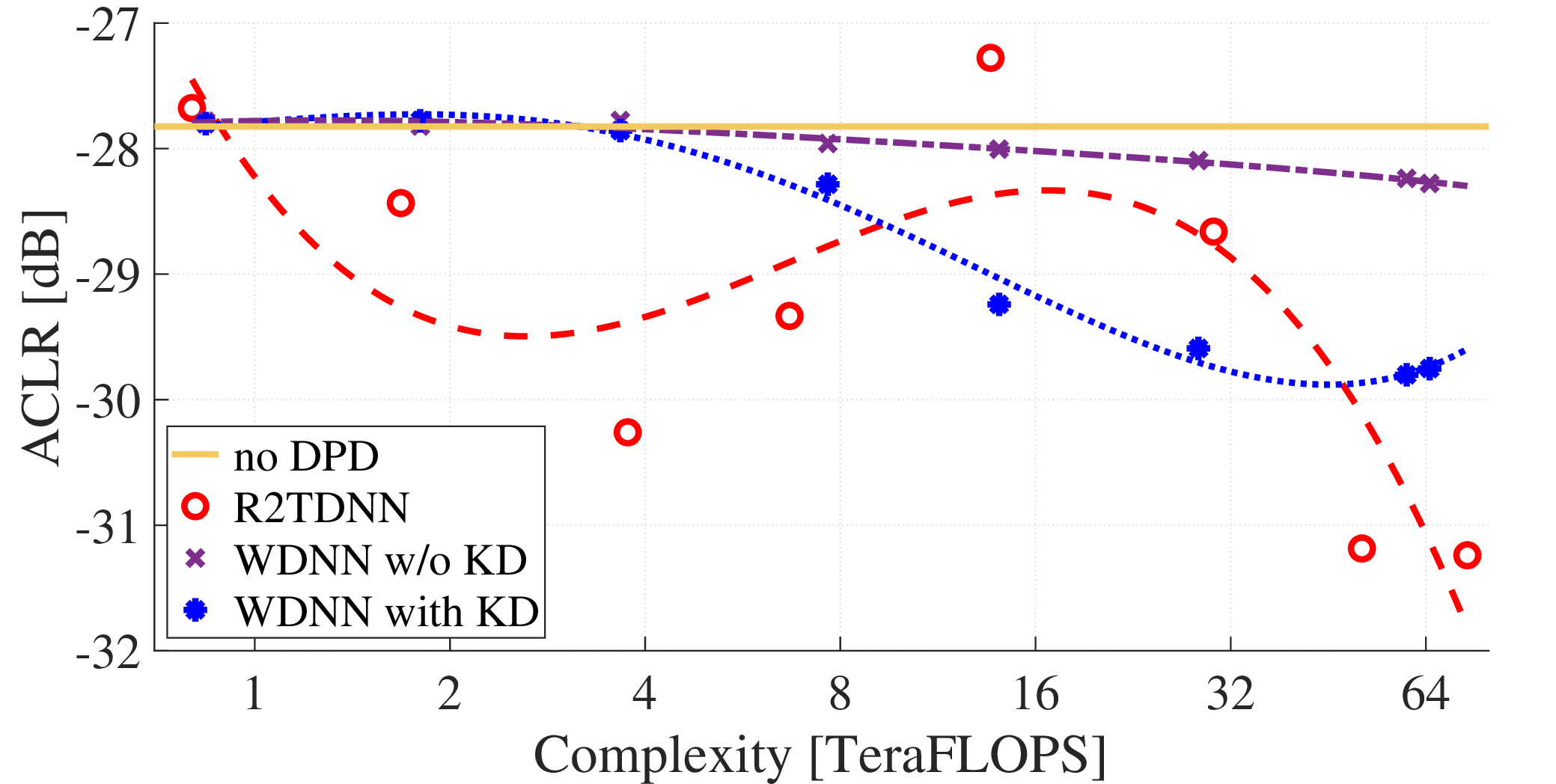}}
\caption{ACLR - complexity tradeoff for investigated DPD methods}
\label{fig:aclrtradeoff}
\vspace{-5mm} 
\end{figure}

Fig.~\ref{fig:spectrumdpd} visualizes how the proposed KD pre-training approach improves the Walsh-based DPD. It improves ACLR by approximately 3dB, achieving comparable linearization performance as IQ-DPD, with a difference of less than 0.08dB in ACLR. We refer to Figs.~\ref{fig:evmtradeoff}~and~\ref{fig:aclrtradeoff} for a quantitative overview of the linearization performance. Here we show EVM \eqref{eq:evm} and ACLR \eqref{eq:aclr} evolution as a function of growing model computational complexity \eqref{eq:flops}. In each figure, the baseline PA performance without linearization is shown as a solid line.

The trend on Fig.~\ref{fig:aclrtradeoff} clearly shows ACLR improvements up to the highest considered complexity, indicating the frequency domain behavior of the PA (the memory effect) requires a higher model complexity. The WDNN-based DPD achieves up to 1.98~dB ACLR improvement after linearization, for model complexity between 16-64~TFLOPS. This is only slightly worse compared to the most complex IQ DPD model. Fig.~\ref{fig:evmtradeoff} illustrates that the EVM for both architectures reaches a comparable level for higher model complexities, with a difference between best models of just 0.24\%. This demonstrates our Walsh-based DPD model reproduces the time-domain features of the ideal Knowledge-Distillation waveform. 

To summarize we show the best achieved PA modelling accuracy and DPD performance in Table~\ref{tab1}. The table shows Walsh-domain neural network predistortion achieves similar performance as the quadrature equivalent. As the DPD performance metrics are similar, our model allows exploiting the benefits of Walsh architecture for reducing the ADC sampling rate and power consumption for ultrawideband communications.

\begin{table}[htbp]
\caption{NN model results summary}
\begin{center}
\begin{tabular}{|c|c|c|c|}
\hline
\textbf{System} & \textbf{NMSE [dB]}& \textbf{ACLR [dB]}& \textbf{EVM [\%]} \\
\hline
PA (no DPD) & / & -27.82 & 8.45 \\
\hline
R2TDNN & -42.31 & -31.24 & 3.36 \\  
\hline
WDNN & -43.03 & -29.80 & 3.12 \\ 
\hline
\end{tabular}
\label{tab1}
\end{center}
\vspace{-5mm} 
\end{table}

\section{Conclusion} \label{sec:conclusion}
An adaptation of TDNN models towards Walsh transceiver architectures was proposed, reducing the power consumption of the analog - digital conversion. An offline pre-training method based on knowledge distillation for domain adaptation improves the spectral mask of the proposed approach and achieves similar linearization performance as traditional quadrature architectures. Simulation results confirm these findings. The proposed adaptation of neural network DPD algorithms paves the way for other signal-processing tasks to be incorporated in the same way into the novel Walsh transceiver architecture.

\ifaccepted
\section*{Acknowledgment}
This work was supported by the European Union’s Horizon 2020 research and innovation programme under grant agreement No 964246. This work only reflects the author's view and the European Commission is not responsible for any use that may be made of the information it contains.
\fi

\bibliography{references}
\bibliographystyle{IEEEtran}

\end{document}